\documentclass[aps,pre,twocolumn,superscriptaddress,showpacs]{revtex4}

\usepackage{bm}
\usepackage{mathrsfs}
\usepackage{graphicx}
\usepackage{color}
\usepackage{amsmath}
\usepackage{amssymb}

\begin{document}

\title{Optimization and universality of Brownian search in quenched heterogeneous
media}

\author{Alja\v{z} Godec}
\email{agodec@uni-potsdam.de}
\affiliation{Institute of Physics \& Astronomy, University of Potsdam, 14776
Potsdam-Golm, Germany}
\affiliation{National Institute of Chemistry, 1000 Ljubljana, Slovenia}
\author{Ralf Metzler}
\email{rmetzler@uni-potsdam.de}
\affiliation{Institute of Physics \& Astronomy, University of Potsdam, 14776
Potsdam-Golm, Germany}
\affiliation{Department of Physics, Tampere University of Technology, FI-33101
Tampere, Finland}

\date{\today}

\begin{abstract}
The kinetics of a variety of transport-controlled processes can be
reduced to the problem of determining the mean time needed to arrive at a
given location for the first time, the so called mean first passage time
(MFPT) problem. The occurrence of occasional large jumps or intermittent
patterns combining various types of motion are known to outperform the
standard random walk with respect to the MFPT, by reducing oversampling of
space. Here we show that a regular but \emph{spatially heterogeneous\/}
random walk can significantly and universally enhance the search in any
spatial dimension. In a generic minimal model we consider a spherically
symmetric system comprising two concentric regions with piece-wise constant
diffusivity. The MFPT is analyzed under the constraint of conserved average
dynamics, that is, the spatially averaged diffusivity is kept constant.
Our analytical calculations and extensive numerical simulations demonstrate
the existence of an {\em optimal heterogeneity} minimizing the MFPT to the
target. We prove that the MFPT for a random walk is completely dominated by
what we term direct trajectories towards the target and reveal a remarkable
universality of the spatially heterogeneous search with respect to target
size and system dimensionality. In contrast to intermittent strategies, which
are most profitable in low spatial dimensions, the spatially inhomogeneous
search performs best in higher dimensions. Discussing our results alongside
recent experiments on single particle tracking in living cells we argue that
the observed spatial heterogeneity may be beneficial for cellular signaling
processes.
\end{abstract}


\maketitle

\section{Introduction}

Random search processes are ubiquitous in nature \cite{Bell,Berg,Lloyd,BrockX,oRMP,
Bray,oNat,Bressloff,Condamin,Holc,Schuss,OlPRL,Tejedor,Amitai,Bress,Gleb3,Gleb4},
ranging from the diffusive motion of regulatory molecules
searching for their targets in living biological cells
\cite{oRMP,Bressloff,Holcman,Sheinmann,Max,Otto,Broek,Mich3,OlPRL2,OlPRL3},
bacteria and animals searching for food by active motion \cite{Berg,Bressloff},
all the way to the spreading of epidemics and pandemics \cite{Lloyd,BrockX}
and computer algorithms in high-dimensional optimization problems
\cite{comps}. The fact that the search strategy in these processes is
to a large extent random reflects the incapability of the searcher
to keep track of his past explorations at least over more than a
certain period \cite{oRMP}. During the years several different search
strategies have been studied in the literature, including Brownian
motion \cite{Sid,Holcman,Bressloff,Ralf}, spatio-temporally decoupled L\'evy
flights (LFs) \cite{Brock1,Brock2,Mich05,Koren,Vladimir,Schehr}
and coupled L\'evy walks (LWs) \cite{ZumKla1,ZumKla2,Geis1,KlaBluShle,KlaShleZu,
Godec1,Godec2,Daniela1,Daniela2,Eli1,Eli2,LWsRMP}
in which the searcher undergoes large re-allocations with a heavy-tailed
length distribution either instantaneously (LFs) or with constant
speed (LWs), as well as intermittent search patterns, in which
the searcher combines different types of motion \cite{oRMP},
for instance, three-dimensional and one-dimensional diffusion
\cite{Holcman,Sheinmann,Max,Otto,Broek,Mich3,OlPRL2,OlPRL3,Mar1,Mar2},
three-dimensional and two-dimensional diffusion
\cite{Ol2D,OlRaph}, or diffusive and ballistic motion
\cite{oRMP,Bressloff,Mich1,Gleb1,Gleb2,ONatPh,LovPRE,BressNew1,BressNew2}.

The efficiency of the search strategy is conventionally quantified via the mean
first passage time (MFPT) defined as the average time a random searcher needs
to arrive at the target for the first time \cite{Sid,Ralf,oRMP,Olivier,Katja}.
The physical principle underlying an improved search efficiency is an optimized
balance between the sampling of space on a scale much larger than the target
and on the scale similar to or smaller than the target \cite{oRMP}.  More
specifically, periods of less-compact exploration---for instance, diffusion
in higher dimensions, L{\'e}vy flights, or ballistic motion---aid towards
bringing the searcher faster into the vicinity of the target. Concurrently
a searcher in such a less compact search mode may thereby easily overshoot
the target \cite{Koren,Vladimir}. In contrast, compact exploration of space
(for instance, diffusion in one dimension) is superior when it comes to
hitting the target from close proximity but performs worse when it comes to
the motion on larger scales taking the searcher from its starting position
into the target's vicinity: typically frequent returns occur to the same
location, a phenomenon referred to as oversampling. Mathematically this is
connected to the recurrent nature of such compact random processes.  The idea
behind search optimization is to find an optimal balance of both more and
less compact search modes in the given physical setting \cite{oRMP}. For
instance in the so-called facilitated diffusion model for the target
search of regulatory proteins on DNA\cite{Sheinmann,Max,Otto,Broek,Mich3}
the average duration of non-compact three-dimensional free diffusion is
balanced with an optimal compact one-dimensional sliding regime along the
DNA molecule. In intermittent search \cite{oRMP,Bressloff,Mich1} persistent
ballistic excursions are balanced by compact Brownian phases. This optimization
principle intuitively works better in lower dimensions, where a searcher
performing a standard random walk oversamples the space. Hence, the typically
considered optimized strategies have the largest gain in low dimensions.

In a variety of experimental situations the motion of a searcher is
characterized by the same search strategy but is not translationally
invariant. A typical example is a system in which the searcher performs a
standard random walk but with a spatially varying rate of making its steps. This
type of motion is actually abundant in biological cells, where experiments
revealed a distinct spatial heterogeneity of the  protein diffusivity
\cite{het1,het2,het4}.  Several aspects of such diffusion in heterogeneous
media have already been addressed \cite{Sid,Thet1,Thet2,Thet3}, but the generic
FPT properties remain elusive, in particular, for quenched environments.

Here was ask the question whether spatial heterogeneity is generically
detrimental for the efficiency of a random search process or whether it could
even be beneficial. Could it even be true that proteins find their targets on
the genome in the nucleus faster because their diffusivity landscape in the
cell is heterogeneous? On the basis of exact results for the MFPT in one, two,
and three dimensions in a closed domain under various settings we here show
that a spatially heterogeneous search can indeed significantly enhance the rate
of arrival at the target. We explain the physical basis of this acceleration
compared to a homogeneous search process and quantify an {\em optimal
  heterogeneity}, which minimizes the MFPT to the target. Furthermore
we show that heterogeneity can be generically beneficial in a random
system and is thus a robust means of enhancing the search kinetics. The
optimal heterogeneous search rests on the remarkable observation that
the MFPT is completely dominated by those trajectories heading directly
towards the target. We prove that the MFPT for the heterogeneous system
can be exactly described with the results of a standard random walk. We
compare our theoretical findings to recent experiments on single particle
tracking in living cells, which are indeed in line with the requirements
for enhanced search.

The paper is organized as follows. Section II introduces our minimal model
for heterogeneous search processes. In Sec.~III we briefly summarize our
main general results, which hold irrespective of the dimension ($d=1$,
$2$, or $3$). Section IV  is devoted to the analysis of the most general
situation with a specific starting point and position of the interface.
In Sec.~V we focus on the Global MFPT, that is, the MFPT averaged over the
initial position. In Section VI we analyze a system with a random position of
the interface and optimize the MFPT averaged over the interface position. In
Section VII we address the Global MFPT in systems with a random position of the
interface. Throughout we discuss our results in a biophysical context motivated
by recent experimental findings. Finally, we conclude by discussing the
implications of our results for more general spatially heterogeneous systems.

\section{Minimal model for spatially heterogeneous random search}    

\begin{figure*}
\includegraphics[width=10cm]{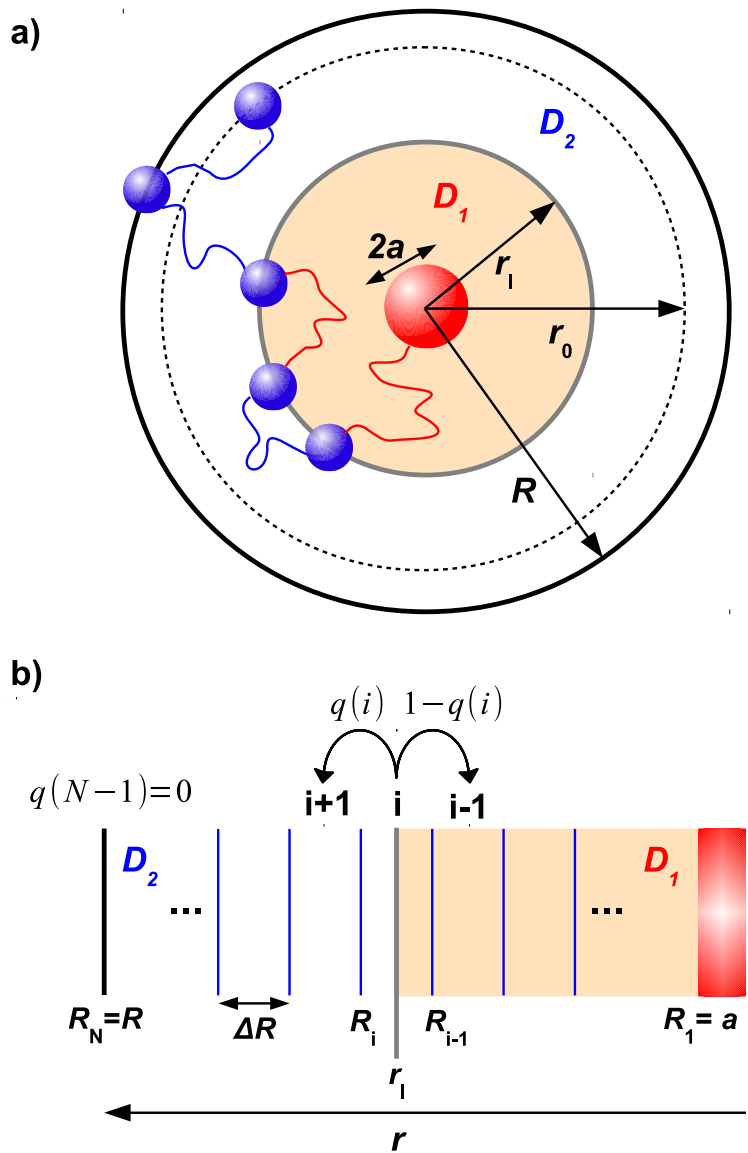}
\caption{Schematic of the model system: a) A spherical target with radius
$a$ is placed in the center of a spherical domain of radius $R$. The
free space between the radii $a$ and $R$ is divided into two regions
denoted by subscripts.  The inner region is bounded by a shell at radius
$r_{\mathrm{I}}$. The outer region ranges from $r_{\mathrm{I}}$ to the
reflective boundary at $R$. Initially, the particle's starting position is
uniformly distributed over the surface of the sphere with radius $r_0$. b)
Microscopic picture of the problem starting from a discrete random walk
between spherical shells. The hopping rates are assumed to obey detailed
balance and the interface position is chosen to be placed symmetrically
between two concentric spherical surfaces.}
\label{schm}
\end{figure*}

We focus on the simplest scenario of a spatially heterogeneous system. Even
for this minimal model the analysis turns out to be challenging and our
exact results reveal a rich behavior with several a priori surprising
features.  We consider a spherically symmetric system in dimensions $d=1$,
$2$, and $3$ with a perfectly absorbing target of radius $a$ located in
the center (Fig.~\ref{schm}). The outer boundary at radius $R$ is taken
to be perfectly reflecting. The system consists of two domains with uniform
diffusivities denoted by $D_1$ and $D_2$ in the interior and exterior domains,
respectively. The interface between these domains is located at $r_\mathrm{I}$.
The microscopic picture we are considering corresponds to the 'kinetic'
interpretation of the Langevin or corresponding Fokker-Planck equations. In
particular we assume that the dynamics obey the  fluctuation-dissipation
relation and in the steady state agree with the results of equilibrium
statistical mechanics \cite{Kinetic}.

In the biological context we consider that the system is in contact with a
heath bath at constant and uniform temperature $T$. The signaling proteins
diffuse in a medium comprising water and numerous other particles, such as
other biomacromolecules or cellular organelles. The remaining particles,
which we briefly call crowders, are not uniformly distributed across the
cell---their identity and relative concentrations differs within the nucleus
and the cytoplasm and can also show variations across the cytoplasm. The
proteins hence experience a spatially varying friction $\Gamma(\bm{r})$,
which originates from spatial variations in the long-range hydrodynamic
coupling to the motion of the crowders, which is in turn mediated by the
solvent \cite{Oppenheim,REM}.  The proteins thus move under the influence of
a position dependent diffusion coefficient $D(\bm{r})=2k_BT\Gamma(\bm{r})$
and a fluctuation-induced thermal drift $\bm{F}(\bm{r})\sim k_BT\nabla\Gamma(
\bm{r})$ (see \cite{Lubensky} for details). The vital role of such hydrodynamic
interactions in the cell cytoplasm was demonstrated in \cite{hydrodyn}.

Because of the spherical symmetry of the problem we can reduce the analysis
to the radial coordinate alone. That is, we only trace the projection of the
motion of the searcher onto the radial coordinate) and therefore start with a
discrete space-time nearest neighbor random walk in-between thin concentric
spherical shells of equal width $\Delta R=R_{i+1}-R_i$ as depicted in
Fig. \ref{schm}.  The shell $i$ denotes the region between surfaces with
radii $R_{i-1}$ and $R_i$. We assume that the hopping rates between shells
$\Pi(i\to j)$ obey detailed balance
\begin{subequations}
\begin{equation}
p(i)\Pi(i\to i+1)=p(i+1)\Pi(i+1\to i).
\label{DB}
\end{equation} 
Here $p(i)$ denotes the probability distribution of finding the particle in
shell $i$. The hopping rates $\Pi$ are given as the product of the intrinsic
rate $2D(i)/\Delta R^2$ and $q(i)$, the probability to jump from shell $i$ to
shell $i+1$ (and $1-q(i)$ for jumps in the other direction),
\begin{equation}
\Pi(i\to i+1)=\frac{2D(i)}{\Delta R^2}q(i).
\label{Hrates}
\end{equation}
Here $D(i)$ is the arithmetic
mean of the diffusivity in shell $i$.
The rate $q(i)$ can be derived as follows. A random
walker located in shell $i$ at time $t$ can either move to shell $i+1$ with
probability $q(i)$ or to shell $i-1$ with probability $1-q(i)$. Due to the
isotropic motion of the random walker these probabilities are proportional to
the respective surface areas of the bounding $d$-dimensional spherical surfaces
at $R_{i-1}$ and $R_i$. That is, $q(i)\sim R_i^{d-1}$ and $1-q(i)\sim
R_{i-1}^{d-1}$. The proportionality constant is readily obtained from the
normalization condition leading to
\begin{equation}
q(i)=\frac{R_i^{d-1}}{R_i^{d-1}+R_{i-1}^{d-1}},
\end{equation}
\end{subequations}
and thus $1-q(i)=R_{i-1}^{d-1}/(R_i^{d-1}+R_{i-1}^{d-1})$. Therefore, while
the random walker moves in all directions (radial, azimuthal, or polar) we
can project its motion on the radial coordinate only.  We assume that the
interface is located between two concentric shells leading to a continuous
steady state probability density profile. The searcher starts at $t=0$
uniformly distributed over the surface of a $d$-sphere with radius $r_0$,
as sketched in Fig.~\ref{schm}a).

In our analytical calculations we model the system in terms of the probability
density function $p(r,t|r_0)$ to find the particle at radius $r$ at time $t$
after starting from radius $r_0$ at $t=0$. $p(r,t|r_0)$ obeys the radial
diffusion equation
\begin{subequations}
\begin{equation}
\label{analytics}
\frac{\partial p(r,t|r_0)}{\partial t}=\frac{1}{r^{d-1}}\frac{\partial }{\partial
r}\left(D(r)r^{d-1}\frac{\partial }{\partial r}\right) p(r,t|r_0)
\end{equation}
with piece-wise constant diffusivity
\begin{equation}
D(r)=\left\{\begin{array}{ll}D_1, & a<r\le r_{\mathrm{I}}\\[0.2cm]
D_2, & r_{\mathrm{I}}<r\le R\end{array}\right..
\end{equation}
We assume that the target surface at radius $a$ is perfectly absorbing,
\begin{equation}
\label{bc_absorb}
p(a,t|r_0)=0,
\end{equation}
to determine the first passage behavior. At the outer radius $R$ we use
the reflecting boundary condition
\begin{equation}
\label{bc_reflect}
\left.\frac{\partial p(r,t|r_0)}{\partial r}\right|_{r=R}=0.
\end{equation}
\end{subequations}
These boundary conditions are complemented with joining conditions at
$r_{\mathrm{I}}$ by requiring the continuity of the probability density and
the flux, which follow from our microscopic picture.

To quantify our model system we introduce the ratio
\begin{subequations}
\begin{equation}
\varphi=\frac{D_1}{D_2}
\end{equation}
of the inner and outer diffusivities. Moreover, we demand that the spatially
averaged diffusivity
\begin{equation}
\label{dcons}
\overline{D}=\frac{d}{R^d-a^d}\int_a^Rr^{d-1}D(r)dr
\end{equation}
\end{subequations}
remains constant for varying $D_1$ and $D_2$. Without such a constraint the
problem of finding an optimal $\varphi$, which minimizes the MFPT is ill-posed
and has a trivial solution $\varphi=\infty$. More importantly, we want to
compare the search efficiency as a function of the degree of heterogeneity,
where the overall intensity of the dynamics is conserved. Returning to our
microscopic picture of a signaling protein searching for its target in the
nucleus, the heterogeneous diffusivity is due to spatial variations in the
long-range hydrodynamic coupling to the motion of the crowders. Their identity
and relative concentration in the cell varies in space, but the effect is
mediated by thermal fluctuations in the solvent at a constant temperature. The
constraint in Eq.~(\ref{dcons}) then corresponds to a redistribution of the
crowders at constant temperature, cell volume and numbers of the various
crowders, which would not affect the spatially averaged diffusivity.

Under the constraint (\ref{dcons}) of constant spatially averaged diffusivity
we obtain for any given $\varphi$ and $r_{\mathrm{I}}$ that
\begin{subequations}
\begin{equation}
D_1=\frac{\varphi\overline{D}}{(\varphi-1)\chi(r_{\mathrm{I}})+1},\,\,\,
D_2=\frac{\overline{D}}{(\varphi-1)\chi(r_{\mathrm{I}})+1},
\label{DC}
\end{equation}
where we introduced the hypervolume ratio
\begin{equation}
\label{cchi}
\chi(r_{\mathrm{I}})=\frac{r_{\mathrm{I}}^d-a^d}{R^d-a^d}.
\end{equation}
\end{subequations}
of the inner versus the entire domain excluding the target volume.
To solve Eq.~(\ref{analytics}) we take a Laplace transform in time and the
obtained Bessel-type equation is solved exactly as shown in Sec.~IV. The MFPT
$\mathrm{\mathbf{T}}(r_0)$ for the particle to reach the target surface at
$r=a$ is obtained from the Laplace transformed flux into the target
\begin{subequations}
\begin{equation}
\label{Gflux}
\widetilde{j}(r_0,s)=\Omega_dD_1a^{d-1}\left.\frac{\partial\widetilde{P}(r,r_0,
s)}{\partial r}\right|_{r=a} 
\end{equation}
via the relation
\begin{equation}
\label{mfpt}
\mathrm{\mathbf{T}}(r_0)=-\frac{\partial \tilde{j}_a(r_0,s)}{\partial
s}\Big|_{s=0}.
\end{equation}
\end{subequations}
The angular prefactor $\Omega_d=1$ for $d=1$, $\Omega_d=2\pi$ for $d=2$ and
$\Omega_d=4\pi$ for $d=3$. We treat the degree of heterogeneity $\varphi$ as an
adjustable parameter at a fixed value of $r_{\mathrm{I}}$ and seek for an optimal
value minimizing the MFPT. The optimal heterogeneity, which we denote with an
asterisk, is thus obtained by extremizing $\mathrm{\mathbf{T}}_a(r_0)$ with
respect to $\varphi$.

\section{Summary of the main results}

\begin{figure*}
\includegraphics[width=12cm]{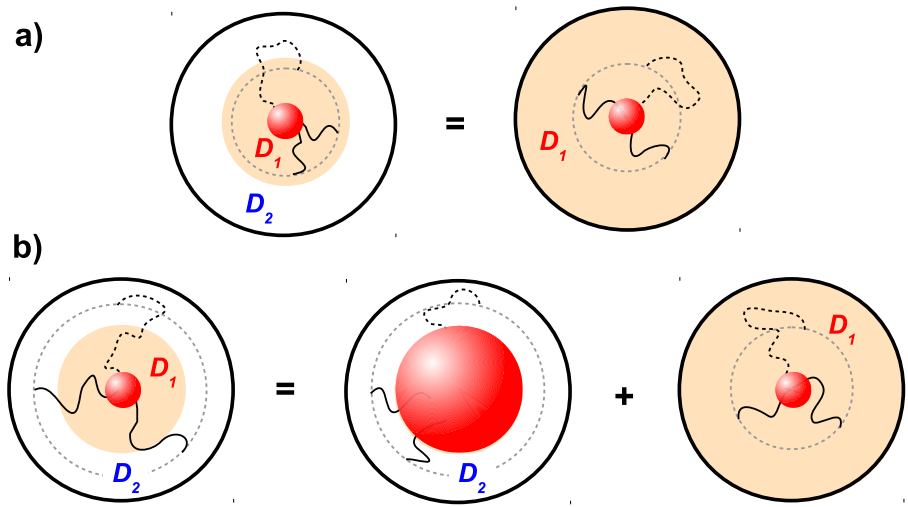}
\caption{Schematic of the equivalence of MFPTs in inhomogeneous and homogeneous
systems in the case of a) a searcher starting in the inner region and b) a searcher
starting in the outer region. The radius $x_1$ of the initial particle position
is shown by the thin dashed circle. We show direct trajectories as full lines.
Each panel also contains an indirect trajectory (dashed line) which leads the
particle into the outer region of the system.}
\label{Sfpt}
\end{figure*}

Since $R$ in combination with the diffusivity $\overline{D}$ sets the
absolute time scale we can express, without loss of generality, time in
units of $R^2/\overline{D}$, set $\overline{D}=1$ and focus on the problem
in a unit sphere. We introduce dimensionless spatial units $x_a=a/R$,
$x_{\mathrm{I}}=r_{\mathrm{I}}/R$ and $x_0=r_0/R$. For the sake of
completeness, we retain the explicit $R$ and $\overline{D}$ dependence in
the prefactors.

Our first main result represents the fact that the MFPT to the target in
the inhomogeneous system in dimension $d=1$, $2$ and $3$, can be expressed
exactly in terms of the corresponding MFPT $\mathrm{\mathbf{T}}^0(x_0;D_i)$
in a homogeneous system with diffusivity $D_i$, with $i=1$ or $2$. Remarkably
the MFPT is thus exactly equal to
\begin{equation}
\label{MFPT}
\mathrm{\mathbf{T}}(x_0)=\left\{\begin{array}{ll}
\mathrm{\mathbf{T}}^0(x_0;D_1), & x_0\le x_{\mathrm{I}}\\[0.2cm]
\mathrm{\mathbf{T}}^0(x_{\mathrm{I}};D_1)+\mathrm{\mathbf{T}}_{x_{\mathrm{I}}}
^0(x_0;D_2), & x_0>x_{\mathrm{I}}\end{array}\right..
\end{equation}
In the second line the argument $x_{\mathrm{I}}$ stands for the release
of the particle at the interface, and the index $x_{\mathrm{I}}$ of the
last term is used to indicate that this term measures the first passage to the
interface at $x_{\mathrm{I}}$. That is, in this case when the particle starts
in the inner region with diffusivity $D_1$ Eq.~(\ref{MFPT}) reveals that the
MFPT of the heterogeneous system is equal to that of a homogeneous system
with diffusivity $D_1$ everywhere and is independent of the position of
the interface. Conversely, if the searcher starts in the exterior region
with diffusivity $D_2$ the MFPT contains two contributions: (i) the MFPT
from $r_0$ to $r_{\mathrm{I}}$ in a homogeneous system with diffusivity
$D_2$ and (ii) the MFPT from $r_{\mathrm{I}}$ to $a$ in a homogeneous
system with diffusivity $D_1$, as shown schematically in Fig.~\ref{Sfpt}.
Eq.~(\ref{MFPT}) is exact and independent of the choice for $D_{1,2}$ and
thus holds for an arbitrary set of diffusivities and even if $D_1=D_2$. In
other words, it is not a consequence of a conserved $\overline{D}$.

The additivity principle of the individual MFPTs in Eq.~(\ref{MFPT})
is only possible if the excursions of the searcher in the directions away
from the target are statistically insignificant. We would expect that some
trajectories starting in the inner region will carry the searcher into the
outer region with diffusivity $D_2$ before they eventually cross the interface
and reach the target by moving through the inner region with diffusivity $D_1$.
Such trajectories will obviously be different in the heterogeneous system
in comparison to a homogeneous system with diffusivity $D_1$ everywhere,
This appears to contradict the complete independence of $D_2$ of the MFPT
in Eq.~(\ref{MFPT}) for trajectories with $x_0\le x_{\mathrm{I}}$. This
observation can be explained by the dominance of \emph{direct trajectories},
whose occupation fraction outside the starting radius is statistically
insignificant: \emph{The MFPT for a standard random walk in dimensions $d=1$,
$2$, and $3$ is completely dominated by direct trajectories}. As such, the MFPT
is really a measure for the efficiency of the fast trajectories.

Our second main result demonstrates that for $x_0>x_{\mathrm{I}}$ a finite optimal
heterogeneity exists at given interface position and is given by
\begin{equation}
\label{optiG}
\varphi^{\ast}(x_0)=\sqrt{\frac{1-\chi(x_{\mathrm{I}})}{\chi(x_{\mathrm{I}})}
\times\frac{\mathbf{T}^0(x_{\mathrm{I}};1)}{\mathbf{T}_{x_{
\mathrm{I}}}^0(x_0;1)}}.
\end{equation}  
For this value the MFPT $\mathrm{\mathbf{T}}(x_0)$ attains a minimum.  Hence,
the optimal heterogeneity is completely determined by the volume fractions
and the MFPT properties and hence strictly by the direct trajectories. As
above, the index $x_{\mathrm{I}}$ in the MFPT $\mathbf{T}^0(x_{\mathrm{I}}$
indicates the first passage to the interface, while without this index the
MFPT $\mathbf{T}^0$ quantifies the first passage to the target at $x_a$. The
explicit results for $\varphi^{\ast}(x_0)$ are shown in Fig.~\ref{OptiMFPT}
and are discussed in detail in Sec.~IV.

Often one is interested in the MFPT averaged over an ensemble of starting
positions, the \emph{Global MFPT} $\overline{\mathrm{\mathbf{T}}}$.
As before, it can be shown that an optimal heterogeneity exists for any
interface position and is universally given by
\begin{eqnarray}
\nonumber
\overline{\varphi}^{\ast}&=&\left(\frac{1-\chi(x_{\mathrm{I}})}{\chi(x_{\mathrm{
I}})}\right.\\
&&\hspace*{-0.8cm}
\times\left.\frac{\mathrm{\mathbf{T}}^0(x_{\mathrm{I}};1)-\int_{x_a}^{x_{
\mathrm{I}}}\left[x_0^d\frac{d}{dx_0}\mathrm{\mathbf{T}}^0(x_0;1)\right]dx_0}{
\mathrm{\mathbf{T}}_{x_{\mathrm{I}}}^0(1;1)-\int_{x_{\mathrm{I}}}^1{\left[x_0^d
\frac{d}{dx_0}\mathrm{\mathbf{T}}_{x_{\mathrm{I}}}^0(x_0;1)\right]dx_0}}\right)
^{1/2}.
\label{GoptiG}
\end{eqnarray}
Similar to the general case $\overline{\varphi}^{\ast}$ is again proportional
to $(\chi(x_{\mathrm{I}})^{-1}-1)^{1/2}$ but here the corresponding MFPTs
in the second factor are reduced by a spatially averaged change of the MFPT
with respect to the starting position. The optimal heterogeneity for the
Global MFPT is shown in Fig.~\ref{FGMFPT}, and discussed in detail in Sec.~V.

In a setting when the interface position is random and uniformly distributed
we are interested in the MFPT from a given starting position averaged over the
interface position. A measurable quantity for this scenario for an ensemble
of random-interface systems is the MFPT $\bm{\{}\mathbf{T}_a(x_0)\bm{\}}$,
where the curly brackets denote an average over the interface positions
$x_{\mathrm{I}}$. Explicit results for dimensions $d=1$, $2$, and $3$ are
given in Sec.~VI. Solving for the optimal heterogeneity we obtain
\begin{widetext}
\begin{equation}
\label{IoptiG}
\bm{\{}\varphi\bm{\}}^{\ast}=\left(\frac{\displaystyle\mathrm{\mathbf{T}}^0(x_0)
-\left(1+\frac{1}{d}\right)\int_{x_a}^{x_0}{x\left(1-\frac{x^d}{d+1}\right)\left[
\frac{d}{dx_{\mathrm{I}}}\mathrm{\mathbf{T}}^0(x_0)\right]dx_{\mathrm{I}}}}
{\displaystyle x_a^{d+1}\mathrm{\mathbf{T}}^0(x_0)-\left(1+\frac{1}{d}\right)\int_
{x_a}^{x_0}x\left(\frac{x^d}{d+1}-x_a^d\right)\left[\frac{d}{dx_{\mathrm{I}}}
\mathrm{\mathbf{T}}_{x_{\mathrm{I}}}^0(x_0)\right]dx_{\mathrm{I}}}\right)^{1/2}.
\end{equation} 
The optimal heterogeneity for the MFPT averaged over the random interface position
is shown in Fig.~\ref{FIMFPT}. 

Finally, we compute the Global MFPT averaged over the position of the interface,
$\bm{\{}\overline{\mathrm{\mathbf{T}}}\bm{\}}$, whose explicit results are given
in Sec.~VI. Also here an optimal strategy can be identified as
\begin{equation}
\label{GGoptiG}
\bm{\{}\overline{\varphi}\bm{\}}^{\ast}=\left(\frac{\displaystyle\overline{\mathrm{
\mathbf{T}}}^0-\int_{x_a}^{1}{(1-x_{\mathrm{I}})x_{\mathrm{I}}^d\frac{1+d-x_{
\mathrm{I}}^d}{1-x_a^d}\frac{d}{dx_{\mathrm{I}}}\mathrm{\mathbf{T}}^0(x_{\mathrm{
I}};1)dx_{\mathrm{I}}}}{\displaystyle x_a^{d+1}\overline{\mathrm{\mathbf{T}}}^0-
\int_{x_a}^{1}{x_{\mathrm{I}}^d\frac{x_{\mathrm{I}}^d-[1+d]x_a^d}{1-x_a^d}\frac{d}{
dx_{\mathrm{I}}}\overline{\mathrm{\mathbf{T}}}_{{x_\mathrm{I}}}^0dx_{x_{\mathrm{I}}}}
}\right)^{1/2}.
\end{equation} 
\end{widetext} 
The optimal heterogeneity for the Global MFPT averaged over the random interface
position is shown in Fig.~\ref{FGGMFPT}. 

Eq.~(\ref{MFPT}) represents a rigorous proof that direct trajectories dominate
the MFPT for Brownian motion. In addition, the heterogeneity does not affect
the fraction of direct versus indirect trajectories but only their respective
durations.  Due to the fact, that indirect trajectories are statistically
insignificant, we can in principle make them arbitrarily slow if we start
in the inner region. But as we let the inner diffusivity go to infinity
(and hence the outer one to zero) we are simultaneously slowing down the
arrivals to the interface if starting from the outer region.  The physical
principle underlying the acceleration of search kinetics is: {\em The optimal
heterogeneity corresponds to an improved balance between the MFPT to reach
the interface and the one to reach the target from the interface.}

\section{Mean first passage time for fixed initial and interface positions}

Here we present the mathematical derivation and the explicit results for the
situation with a specific initial condition $r_0$ and interface position $r_{
\mathrm{I}}$. Eq.~(\ref{analytics}) is solved by Laplace transformation, and
the resulting Green's function with the boundary conditions (\ref{bc_absorb})
and (\ref{bc_reflect}) reads
\begin{subequations}
\begin{widetext}
\begin{equation}
\label{Green}
\widetilde{P}(r,s|r_0)={\frac{(rr_0)^{\nu}}{\Omega_dD_1}\frac{\mathcal{C}_{\nu}(
S_1r,S_1a)}{\displaystyle\frac{\mathcal{D}_{\nu}(S_1a,S_1r_{\mathrm{I}})}{
\mathcal{C}_{\nu-1}(S_2r_{\mathrm{I}},S_2R)}+\frac{1}{\sqrt{\varphi}}\frac{
\mathcal{C}_{\nu}(S_1a,S_1r_{\mathrm{I}})}{\mathcal{D}_{\nu}(S_2r_{\mathrm{I}}
,S_2R)}}}\left\{\begin{array}{ll}\displaystyle{\frac{\mathcal{D}_{\nu}(S_1r_0,
S_1r_{\mathrm{I}})}{\mathcal{C}_{\nu-1}(S_2r_{\mathrm{I}},S_2R)}+\frac{1}{\sqrt{
\varphi}}\frac{\mathcal{C}_{\nu}(S_1r_0,S_1r_{\mathrm{I}})}{\mathcal{D}_{\nu}(S_2
r_{\mathrm{I}},S_2R)}}, & a<r_0\le r_{\mathrm{I}}\\
\displaystyle{\frac{\mathcal{D}_{\nu}(S_2r_0,S_2R)}{r_{\mathrm{I}}S_1\mathcal{D}_{
\nu}(S_2r_{\mathrm{I}},S_2R)\mathcal{C}_{\nu-1}(S_2r_{\mathrm{I}},S_2R)}},
& r_{\mathrm{I}}<r_0\le R
\end{array}\right.,
\end{equation}
\end{widetext}
where we introduced the abbreviation $S_{1,2}=\sqrt{s/D_{1,2}}$ and the auxiliary
functions
\begin{eqnarray}
\nonumber
\mathcal{D}_{\nu}(z_1,z_2)&=&I_{\nu}(z_1)K_{\nu-1}(z_2)+K_{\nu}(z_1)I_{\nu-1}(z_2)\\
\mathcal{C}_{\nu}(z_1,z_2)&=&I_{\nu}(z_1)K_{\nu}(z_2)-I_{\nu}(z_2)K_{\nu}(z_1).
\label{Aux2}
\end{eqnarray}
\end{subequations}
Here the $I_{\nu}(z)$ and $K_{\nu}(z)$ denote the modified Bessel functions
of order $\nu=1-d/2$ of the first and second kind, respectively. The Laplace
transformed first passage time density is obtained from the flux (\ref{Gflux})
into the target, and the MFPT then follows from relation (\ref{mfpt}).
In case of $d=1$ the target size only enters the problem by determining
the width of the interval. Using $\mathcal{D}_{\nu}(z,z)=1/z$ it can be
shown that Eq.~(\ref{Green}) reduces to the ordinary expression for regular
diffusion given in Ref. \cite{Sid} for $r_{\mathrm{I}}=R$ and $\varphi=1$.

\begin{figure*}
\includegraphics[width=18.cm]{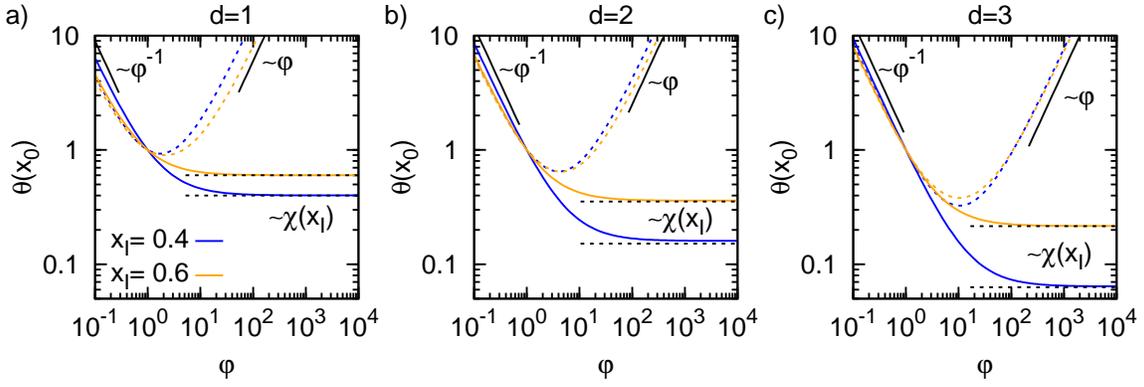}
\caption{Ratio
$\theta(x_0)=\mathrm{\mathbf{T}}(x_0)/\mathrm{\mathbf{T}}^0(x_0)$ as function
of $\varphi=D_1/D_2$ in dimension a) $d=1$, b) $d=2$, and c) $d=3$ for initial
radius $x_0=0.3$ (dashed lines) and $x_0=0.8$ (full lines), respectively. In
b) and c) we plotted the results for a target size $x_a=0.1$. The full black
lines denote the $\simeq 1/\varphi$ and $\simeq \varphi$ scaling, respectively,
and the dashed black line corresponds to $\simeq\chi(x_{\mathrm{I}})$.}
\label{RMFPT}
\end{figure*}

The MFPT can be written exactly in terms of the expressions for a homogeneous
diffusion process in Eq.~(\ref{MFPT}) (compare Refs.~\cite{Sid,Olivier}), and
we obtain
\begin{subequations} 
\label{GNRL}
\begin{eqnarray}
\label{MFPT1}
_1\mathrm{\mathbf{T}}^0(x_0;\overline{D})&=&\frac{R^2}{2\overline{D}}
[2(x_0-x_a)+x_a^2-x_0^2],\\
\label{MFPT2}
_2\mathrm{\mathbf{T}}^0(x_0;\overline{D})&=&\frac{R^2}{4\overline{D}}
\left[2\log\left(\frac{x_0}{
x_a}\right)+x_a^2-x_0^2\right],\\
\label{MFPT3}
_3\mathrm{\mathbf{T}}^0(x_0;\overline{D})&=&\frac{R^2}{6\overline{D}}
\left[2\frac{x_0-x_a}{x_0x_a}+x_a^2-x_0^2\right].
\end{eqnarray} 
\end{subequations}
Here the left index denotes the dimensionality of the system.
Our results show excellent agreement with numerical
simulations, as demonstrated in the Appendix. Plugging the above expressions
(\ref{GNRL}) into Eq.~(\ref{MFPT}) we can compare
the search efficiency with respect to a homogeneous random walk by
introducing the dimensionless ratio
\begin{equation}
\theta(x_0)=\frac{\mathrm{\mathbf{T}}(x_0)}{\mathrm{\mathbf{T}}^0(x_0)},
\end{equation}
The corresponding results for dimensions $d=1$, 2, and 3 are shown in
Fig.~\ref{RMFPT}. 

The qualitative behavior of the MFPT with respect to $\varphi$---that
is, the degree of the heterogeneity---depends on the starting position
relative to the interface. If the initial position lies in the inner region
$\theta(x_0)$ decays monotonically and saturates at a finite asymptotic
value, $\theta(x_0)\to\chi(x_{\mathrm{I}})$. Hence, in this case an optimal
strategy does not exists and the best search performance is achieved for
large diffusivities in the inner region. The lower bound on $\theta(x_0)$
is set by the volume fraction of the inner region, which sets a bound on
the ratio $D_1/\overline{D}$. This result is yet another manifestation of
the fact that the MFPT is completely dominated by direct trajectories.

Conversely, if the searcher starts in the outer region an optimal strategy
exists. To understand the existence and meaning of the optimal heterogeneity
we perform a power series expansion of $\theta(x_0)$. We find the scaling
$\theta(x_0)\simeq 1/\varphi$ as $\varphi \to 0$, which is due to the fact
that in this regime $D_1 \simeq \varphi$, which dominates the MFPT in this
regime. In the other limit as  $\varphi \to \infty$ we find $\theta(x_0)\simeq
\varphi$ because here $D_2\simeq \varphi$ and the rate determining step is
the arrival at the interface. We can understand the optimal heterogeneity
as a beneficial balance between the rate of arriving at the interface from
the starting position and the rate to find the target if starting from
the interface.

\begin{figure*}
\includegraphics[width=18.cm]{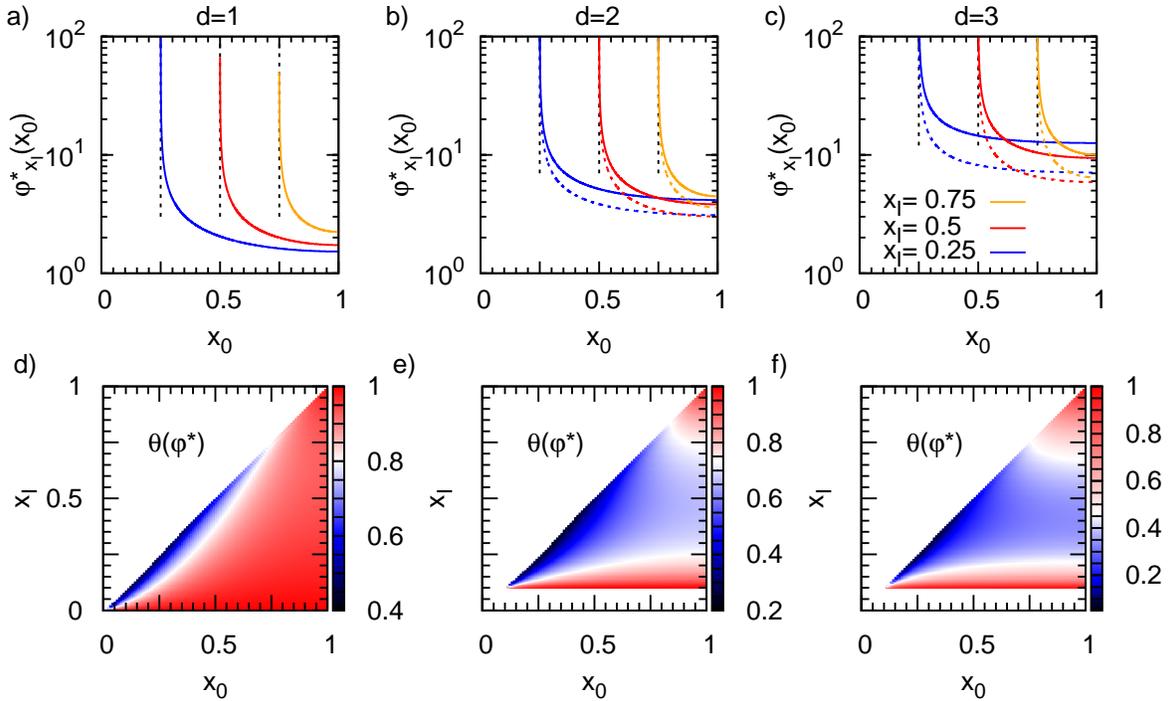}
\caption{Optimal heterogeneity as function of the starting position
$x_0$ for three interface positions $x_{\mathrm{I}}=0.25$ (blue),
$x_{\mathrm{I}}=0.5$ (red), and $x_{\mathrm{I}}=0.75$ (orange) in
dimension a) $d=1$, b) $d=2$, and c) $d=3$. The target sizes in b)
and c) are $x_a=$0.1 (full lines) and $x_a=$0.2 (dashed lines). Ratio
$\theta(x_0)=\mathrm{\mathbf{T}}(x_0)/\mathrm{\mathbf{T}}^0(x_0)$ for the
optimal heterogeneity $\varphi^{\ast}$ in dimension d) $d=1$, e) $d=2$,
and f) $d=3$. In e) and f) the target size is $x_a=0.1$.}
\label{OptiMFPT}
\end{figure*}

In contrast, too large values of $\varphi$ prolong the time to reach
the interface and cannot be compensated by a faster arrival from the
interface towards the target.  The optimal heterogeneity as a function
of the starting position is shown in Fig.~\ref{OptiMFPT}a)-c) and reveals
the divergence as $x_0\to x_{\mathrm{I}}$: this point corresponds to the
disappearance of an optimal strategy. As $x_0$ gradually approaches unity,
$\varphi^{\ast}$ continuously decreases towards a plateau, which suggests
that to reach the target from the interface becomes the rate determining
step. Moreover, $\varphi^{\ast}$ decreases with increasing target size,
because the inner region becomes smaller and thus allows a larger $D_2$ in
the optimal scenario.  The overall gain of an optimal search with respect
to a standard random walk is shown in Fig.~\ref{OptiMFPT}d)-f). Only in the
domain $x_0>x_{\mathrm{I}}$ an optimal heterogeneity exists, therefore we
omitted the region $x_0<x_{\mathrm{I}}$. As mentioned before, we observe
the monotone convergence $\theta(x_0)\to\chi(x_{\mathrm{I}})$ from above as
$\varphi\to\infty$ and hence the heterogeneous search always outperforms the
standard Brownian search for $\varphi>1$. The highest gain is therefore set
by the volume fraction of the inner region which becomes arbitrarily small as
$x_{\mathrm{I}}\to x_a$. For $x_0>x_{\mathrm{I}}$ we find that the gain is
largest for starting positions near the interface and when the interface is
not too close to the outer boundary, where the system essentially approaches
the homogeneous limit. The variations are larger in higher dimensions, which
is, of course, connected with the pronounced spatial oversampling in lower $d$.

Different from conventional strategies---intermittent or
L\'evy-stable processes, which affect the compactness of exploring
space---heterogeneous search is more profitable in higher dimensions (see
Fig.~\ref{OptiMFPT}d)-f)). Because heterogeneity acts by enhancing/retarding
the local dynamics and does not affect spatial oversampling it is intuitive
that it performs better for non-compact exploration of space. Both the
existence as well as the gain of an optimally heterogeneous search are thus
a direct consequence of direct trajectories dominating the MFPT.

\section{Global mean first passage time for fixed interface position}

\begin{figure*}
\includegraphics[width=18.cm]{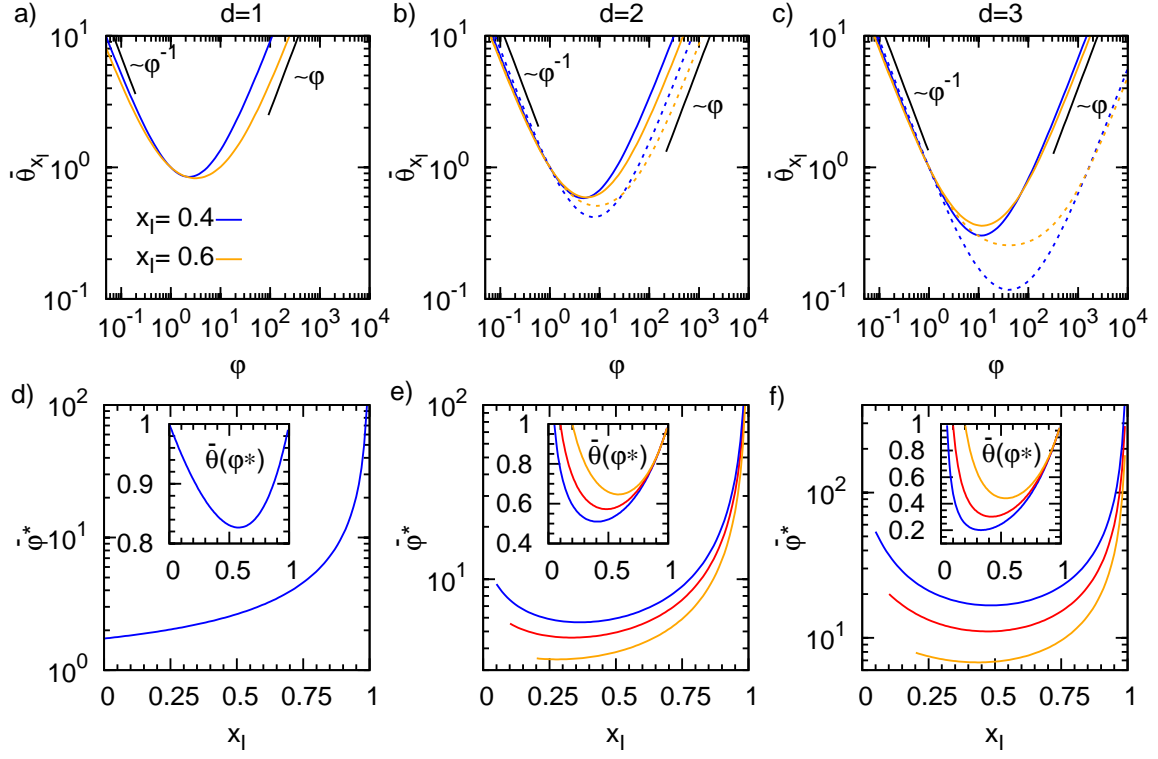}
\caption{Ratio $\overline{\theta}=\overline{\mathbf{T}}/\overline{\mathbf{T}}^0$
as function of $\varphi=D_1/D_2$ for different dimensions and interface positions
$x_{\mathrm{I}}=0.4$ (blue) and $x_{\mathrm{I}}=0.6$ (orange). The target size in
panels b) and c) is $x_a=0.1$ (full lines) and $x_a=0.2$ (dashed lines). d)-f)
Optimal heterogeneity for different dimensions as a function of the interface
position $x_{\mathrm{I}}$. In panels e) and f) the target size is $x_a=$0.05 (blue),
$x_a=$0.1 (red), and $x_a=$0.2 (orange). Insets in panels d)-f):  $\overline{\theta}
$ for the optimal heterogeneity $\overline{\varphi}^{\ast}$.}
\label{FGMFPT}
\end{figure*}

The Global MFPT is obtained by direct averaging of Eq.~(\ref{MFPT}) over
the initial position $x_0$,
\begin{equation}
\overline{\mathbf{T}}=\frac{d}{1-x_a^d}\int_{x_a}^1x_0^{d-1}\mathbf{T}(x_0)dx_0.
\end{equation}
The exact expressions for the Global MFPT in the various dimensions read
\begin{widetext}
\begin{subequations}
\begin{eqnarray}
\label{GMFPT1}
_{1}\overline{\mathrm{\mathbf{T}}}&=&\frac{\displaystyle R^2\left(\left[1-
\frac{1}{\varphi}\right]x_{\mathrm{I}}+\frac{1}{\varphi}\right)}{\overline{D}}
\left[\frac{\varphi}{3}+x_{\mathrm{I}}(1-\varphi)-x_{\mathrm{I}}^2(1-\varphi)+
\frac{x_{\mathrm{I}}^3(1-\varphi)}{3}\right],\\
\nonumber
_{2}\overline{\mathrm{\mathbf{T}}}&=&\frac{\displaystyle R^2\left(\left[1-\frac{1}{
\varphi}\right]x_{\mathrm{I}}^2+\frac{1}{\varphi}-x_a^2\right)}{4\overline{D}(1-x
_a^2)^2}\\
\label{GMFPT2}
&&\times\left[2(1-\varphi)\log(x_{\mathrm{I}})-2\log(x_a)-(x_{\mathrm{I}}^2-x_a^2)
\left(2-\frac{x_{\mathrm{I}}^2+x_a^2}{2}\right)-\varphi\left(\frac{3}{2}-2x_{\mathrm{I}}+
\frac{x_{\mathrm{I}}^4}{2}\right)\right],\\[0.2cm]
\label{GMFPT3}
_{3}\overline{\mathrm{\mathbf{T}}}&=&\frac{\displaystyle R^2\left(\left[1-\frac{
1}{\varphi}\right]x_{\mathrm{I}}^3+\frac{1}{\varphi}-x_a^3\right)}{6\overline{D}(1-x_a^3)^2}\left[\frac{2}{x_a}+\frac{2(\varphi-1)}{x_{\mathrm{I}}}+2x_{\mathrm{I}}^2(\varphi-1)\left(1-\frac{x_{\mathrm{I}}^3}{5}\right)-\frac{9\varphi}{5}+2x_a^2\left(1-\frac{x_a^3}{5}\right)\right].
\end{eqnarray} 
\end{subequations}
\end{widetext}
These are to be compared with their homogeneous counterparts
\begin{subequations}
\begin{eqnarray}
\label{hGMFPT1}
_{1}\overline{\mathrm{\mathbf{T}}}^0&=&\frac{R^2}{3\overline{D}},\\
\nonumber
_{2}\overline{\mathrm{\mathbf{T}}}^0&=&\frac{R^2}{4\overline{D}(1-x_a^2)^2}\\
\label{hGMFPT2}
&&\times\left[-\frac{3}{2}+2x_a^2-\frac{x_a^4}{2}-2\log(x_a)\right],\\
\nonumber
_{3}\overline{\mathrm{\mathbf{T}}}^0&=&\frac{R^2}{3\overline{D}(1-x_a^3)^2}\\
\label{hGMFPT3}
&&\times\left[1-\frac{9x_a}{5}+x_a^3\left(1-\frac{x_a^3}{5}\right)\right],
\end{eqnarray}
\end{subequations}
in $d=1$, $2$, and $3$, respectively. As before we introduce the dimensionless
enhancement ratio
\begin{equation}
\overline{\theta}=\frac{\overline{\mathrm{\mathbf{T}}}}{\overline{\mathrm{
\mathbf{T}}}^0}.
\end{equation}
The result are shown in Fig.~\ref{FGMFPT}.  In this case we are effectively
considering a weighted average of the results presented in the previous
Sec. IV. Noticing that most of the volume of a $d$-sphere is increasingly
concentrated near the surface for growing dimensions we anticipate that the
results will be more prominently influenced by the features of trajectories
starting further away from the target. An interesting question will therefore
be whether an optimal heterogeneity exists for all interface positions and
dimensions. An expansion of $\overline{\mathrm{\mathbf{T}}}$ in a power
series of $\varphi$ reveals that $\overline{\theta}\simeq 1/\varphi$ as
$\varphi\to 0$ and $\overline{\theta}\simeq \varphi$ as $\varphi\to \infty$
for all positions of the interface (see Fig.~\ref{FGMFPT} a)-c)). Thus, an
optimal heterogeneity always exists for all $d$.  From  Fig.~\ref{FGMFPT}
a)-c) we observed that the dependence of the relative gain compared
to the $\overline{\theta}$ on the interface position in the limit
$\varphi\to 0$ is very weak in all dimensions as long as it is not too
close to the external boundary. In the case of large $\varphi$ somewhat
larger variations are observed for dimensions $d=1$ and 2. In $d=3$ the
dependence on the interface position is largest near the optimal value
for $\overline{\varphi}$ but very weak elsewhere. Hence, a control of the
target location dynamics by adjusting $x_{\mathrm{I}}$ is only efficient near
the optimal point $\overline{\varphi}=\overline{\varphi}^{\ast}$.  This is
observed from the respective scaling $\overline{\theta}\simeq 1/\varphi$ and
$\overline{\theta}\simeq \varphi$ as $\varphi\to 0$ and $\varphi\to \infty$
as explained in Sec.~IV.  Overall the gain with respect to the homogeneous
random walk is larger for higher dimensions, which has the same origin as
in the general case discussed above, however here the additional effect of
averaging over the initial position enters.  The optimal heterogeneity has
the exact results
\begin{widetext}
\begin{subequations} 
\label{GGOopti}
\begin{eqnarray}
\label{optiG1}
_{1}\overline{\varphi}^{\ast}&=&\frac{\sqrt{3(1-x_{\mathrm{I}})+x_{\mathrm{I}}
^2}}{1-x_{\mathrm{I}}},\\
\label{optiG2}
_{2}\overline{\varphi}^{\ast}&=&\left(\frac{\displaystyle(1-x_{\mathrm{I}}^2)\left[
-4+x_{\mathrm{I}}^2+x_a^2+\frac{4}{x_{\mathrm{I}}^2-x_a^2}\log\left(\frac{x_{
\mathrm{I}}}{x_a}\right)
\right]}{-4\log(x_{\mathrm{I}})-3+4x_{\mathrm{I}}^2-x_{\mathrm{I}}^4}\right)^{1/2},\\
_{3}\overline{\varphi}^{\ast}&=&\left(\frac{1-\chi(x_{\mathrm{I}})}{\chi(x_{\mathrm{
I}})}\frac{5[\frac{x_{\mathrm{I}}}{x_a}(1+x_a^3)-(1+x_{\mathrm{I}}^3)]+x_{\mathrm{
I}}(x_{\mathrm{I}}^5-x_a^5)}{(1-x_{\mathrm{I}})^3(5+x_{\mathrm{I}}[6+x_{\mathrm{I}}
(3+x_{\mathrm{I}})])}\right)^{1/2}.
\label{optiG3}
\end{eqnarray} 
\end{subequations}
\end{widetext}
The behavior of Eqs.~(\ref{GGOopti}) is shown in Fig.~\ref{FGMFPT}e)-f).
In higher dimensions the optimal heterogeneity shows a non-monotonic behavior
with respect to the interface position and increases upon approaching the
target or the outer boundary. Simultaneously, the overall dependence of
the Global MFPT on $\varphi$ vanishes in these limits---see the inset of
Fig.~\ref{FGMFPT})e)-f)---corresponding to the situation when the system
is no longer heterogeneous. These results can be rationalized by the fact
that in the limit $x_{\mathrm{I}}\to x_a$ the ratio $\chi(x_{\mathrm{I}})$
becomes negligible and hence a higher $D_1$ is allowed without slowing down
the dynamics of reaching the interface from the external region. Conversely,
as $x_{\mathrm{I}}\to 1$ also $\chi(x_{\mathrm{I}})\to 1$ and smaller  $D_2$
are allowed because the rate limiting step is hitting the target from the
inner region. In both limits, however, the overall enhancement effect with
respect to a standard random walk becomes negligible.  Away from these limits
the gain of the optimal heterogeneity is larger for higher dimensions and
can be remarkably large for intermediate interface positions (see
the inset of Fig.~\ref{FGMFPT})e)-f) and increases with decreasing target size.

In a biological context, the present setting is relevant for signaling proteins
searching for their target in the nucleus when the proteins are initially
uniformly distributed throughout the cell cytoplasm. Recent experiments
revealed a significant heterogeneity in the spatial dependence of the protein
diffusion coefficient across the cell, with a faster diffusivity near the
nucleus \cite{het1,het2}. Such a spatial heterogeneity could therefore be
beneficial for the cell by accelerating the dynamics of signaling molecules.

\section{Mean first passage time in a random heterogeneous system for fixed
initial position}

\begin{figure*}
\includegraphics[width=18.cm]{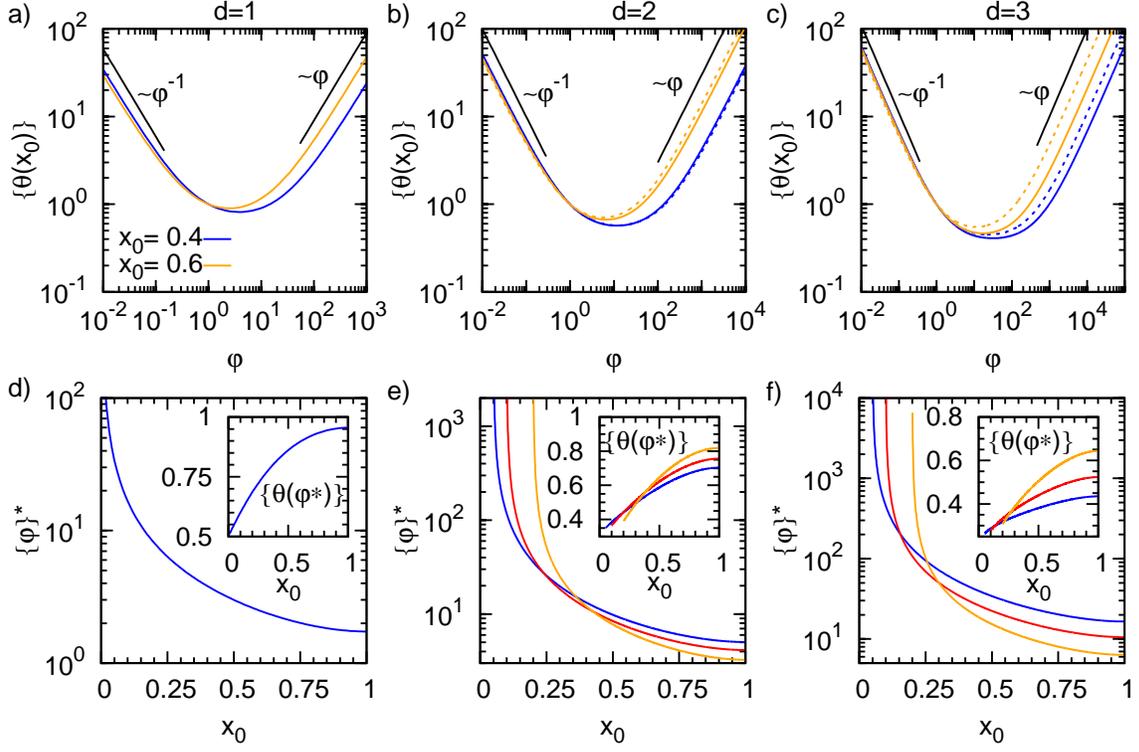}
\caption{Ratio $\bm{\{}\theta\bm{\}}=\bm{\{}\mathbf{T}\bm{\}}/\mathbf{T}(x_0)$
as function of $\varphi=D_1/D_2$ for different dimensions and  $x_0=0.4$
(blue) and $x_0=0.6$ (orange). The target size in panels b) and c) is $x_a=0.1$
(full lines) and $x_a=0.2$ (dashed lines). d)-f) Optimal heterogeneity for
different dimensions as a function of the starting position $x_0$. In panels e)
and f) the target size is $x_a=0.05$ (blue), $x_a=0.1$ (red), and $x_a=0.2$
(orange). Inset in panels d)-f): $\bm{\{}\theta\bm{\}}$ for the optimal
heterogeneity $\bm{\{}\varphi^{\ast}\bm{\}}$.}
\label{FIMFPT}
\end{figure*}

We now address the MFPT problem when the interface position is random
in a given realization and uniformly distributed over the radial
domain. Specifically, we are interested in the MFPT of a particle starting
at $x_0$ averaged over the interface position $x_{\mathrm{I}}$,
\begin{equation}
\mathbf{\bm{\{}}\mathbf{T}\bm{\}}(x_0)=\frac{1}{1-x_a}\int_{x_a}^{1} 
{\mathrm{\mathbf{T}}}(x_0)dx_{\mathrm{I}}.
\end{equation}
Experimentally, this would correspond to measuring an ensemble of systems with
random value of $x_{\mathrm{I}}$. For dimensions $d=1$, $2$, and $3$ the MFPT
$\bm{\{}\mathbf{T}\bm{\}}(x_0)$ has the explicit form  
\begin{widetext}
\begin{subequations}
\label{AP1X} 
\begin{eqnarray}
\label{IMFPT1}
_1\bm{\{}\mathbf{T}\bm{\}}(x_0)&=&\frac{R^2}{2\overline{D}}x_0\left[1+\frac{1}{
\varphi}+\left(1-\frac{3}{\varphi}\right)x_0\left(\frac{1}{2}+\left(1-\frac{1}{
\varphi}\right)x_0\right)+\frac{(1-\varphi)^2x_0^3}{4\varphi}\right],\\[0.2cm]
\nonumber
_2\bm{\{}\mathbf{T}\bm{\}}(x_0)&=&\frac{R^2}{12\overline{D}(1-x_a^2)(1-x_a)}
\left[\left(1+\frac{1}{\varphi}\right)x_0\left\{6(1-\varphi x_a^2)-\frac{2x_0^2[
4-\varphi(1+3x_a^2)]}{3}-\frac{2x_0^4(\varphi-1)}{5}\right\}\right.\\
\nonumber
&&-\left\{\frac{2}{\varphi}+1-x_a(3+x_a[3-x_a\{1+2\varphi\}])\right\}\left[2\log
\left(\frac{x_0}{x_a}\right)-x_0^2\right]-\left\{6\left(1-\frac{1}{\varphi}\right)
\right.\\
\label{IMFPT2}
&&\left.\left.-x_a\left[1+\displaystyle{\frac{6-x_a[8-\varphi(16\varphi-17)]}{3\varphi}}\right]
-x_a^2\left[3-x_a^2\frac{2+\varphi\{11+2\varphi\}}{\varphi}\right]\right\}
\right].\\[0.2cm]
\nonumber
_3\bm{\{}\mathbf{T}\bm{\}}(x_0)&=&\frac{R^2}{8\overline{D}(1-x_a)(1-x_a^3)}\left[
\frac{1}{3}\left(\frac{1}{\varphi}-1\right)x_0\left[2(5-\varphi\{1-4x_a^3\})-(1-
\varphi)x_0^3\right]\right.\\
\nonumber
&&+\frac{x_a^3}{3}\left[2(4\varphi-5\frac{1}{\varphi})+x_a^3\left(\frac{1}{\varphi}
+2\varphi\right)\right]+\left(\frac{2}{x_a}+x_a^2\right)\left[\frac{3}{\varphi}+1
-4x_a\left(1+x_a^2\left\{1-\frac{3x_a}{4}\right\}\right)\right]\\
\nonumber
&&-\left(\frac{2}{x_0}+x_0^2\right)\left[\frac{3}{\varphi}+1-4x_a\left(1+x_a^2
\left\{1-\frac{(3\varphi+1)x_a}{4}\right\}\right)\right]\\
\label{IMFPT3}
&&\left.-8\left(\frac{1}{\varphi}-1\right)(1-\varphi x_a^3)\log\left(\frac{x_0}{x
_a}\right)\right].
\end{eqnarray} 
\end{subequations}
\end{widetext}
The gain compared to the homogeneous random walk,
\begin{equation}
\bm{\{}\theta(x_0)\bm{\}}=\frac{\bm{\{}\mathbf{T}\bm{\}}(x_0)}{\mathbf{T}(x_0)},
\end{equation}
is shown in Fig.~\ref{FIMFPT}a)-c) and reveals a somewhat less sharp optimal
heterogeneity as compared to the Global MFPT. The limiting behavior for small
and large $\varphi$ is the same as for the previous cases but here there is a
wider range of $\varphi$ values producing a comparable gain. We also observe
a stronger dependence on the starting position as $\varphi \to \infty$ and a
very weak dependence for $\varphi \to 0$, which is yet another consequence
of direct trajectories dominating the MFPT. The optimal heterogeneity is
obtained from Eq.~(\ref{IoptiG}) and reads
\begin{widetext}
\begin{subequations} 
\begin{eqnarray}
\label{optiI1}
_1\bm{\{}\varphi\bm{\}}^{\ast}&=&\frac{1}{x_0}\left(\frac{3(2-x_0)(2-x_0[2-x_0])}{
4-3x_0}\right)^{1/2},\\
\label{optiI2}
_2\bm{\{}\varphi\bm{\}}^{\ast}&=&\left(\frac{2\log(x_0/x_a)-3[Q_1(x_0)-Q_1(x_a)]}{
2x_a^3\log(x_0/x_a)+Q_2(x_0,x_a)}\right)^{1/2},\\
\label{optiI3}
_3\bm{\{}\varphi\bm{\}}^{\ast}&=&\left(\frac{24\log(x_0/x_a)+Q_3(x_0)-Q_3(x_a)}{
24x_a^3\log(x_0/x_a)-Q_4(x_0,x_a)}\right)^{1/2}.
\end{eqnarray} 
\end{subequations}
Here we introduced the auxiliary functions
\begin{subequations} 
\label{AUXX}
\begin{eqnarray}
Q_1(y)&=&y\left(3+y\left[1-\frac{4y}{3}+\frac{y^3}{5}\right]\right),\\
Q_2(y,z)&=&y^3\left(\frac{1}{3}-\frac{y^2}{5}\right)-z^2\Big(y[3-y^2]
\left.-z\left[\frac{8}{3}-y^2\right]-\frac{z^3}{5}\right),\\
Q_3(y)&=&\frac{18}{y}+y^2(9-y[10-y^3]),\\
Q_4(y,z)&=&y^3(2[1+y^2z]-y^3)+z^3(2+y^3)\left(8-\frac{9z}{y}\right).
\end{eqnarray} 
\end{subequations}
\end{widetext}

The results for various dimensions are depicted in
Fig.~\ref{FIMFPT}d)-\ref{FIMFPT}f). Accordingly an optimal heterogeneity
always exists. Note that the dependence of $\bm{\{} \varphi \bm{\}}^{\ast}$
on the target size in $d=2$ and 3 becomes reversed upon increasing the initial
separation to the target. When starting near the target the initial position
will on average lie in the inner region in an ensemble of realizations
of the interface position. The optimal heterogeneity will thus on average
correspond to a very fast diffusion in the inner region. For small $x_0$
the probability of starting in the inner region, $(1-x_0)/(1-x_a)$, will be
lower for smaller targets and hence $\bm{\{} \varphi \bm{\}}^{\ast}$ will
be larger accordingly. Conversely, if starting further away from the target
$x_0$ will on average lie in the outer region and the search time will be
more strongly influenced by the rate of arriving at the interface in each
realization. An optimal heterogeneity will therefore correspond to a smaller
asymmetry of diffusivities in the inner and outer regions. More specifically,
since the probability of starting in the outer region, $(x_0-x_a)/(1-x_a)$,
will be lower for larger targets the value of $\bm{\{} \varphi \bm{\}}^{\ast}$
will accordingly be smaller. The gain of the optimal heterogeneity is shown
in the insets of Fig.~\ref{FIMFPT}d)-\ref{FIMFPT}f) and reveals a significant
improvement with respect to the standard random walk in every dimension.
The dependence on target size has again a non-monotonic behavior, which
follows from the argument presented above. Hence, even in a system with
a random and quenched position of the interface a spatially heterogeneous
search process will on average strongly outperform the standard random walk
in every dimension.

In a biological context the present setting is important for the experimentally
observed quenched spatial disorder revealed by single-particle tracking
\cite{het4,Thet3,het3,het2,Mhet2,Mhet3}. More generally, the ideas can also
be extended to the dynamics in complex disordered systems \cite{dis1,dis2}
and Sinai-type diffusion \cite{Sinai}.  Our results show that quenched spatial
heterogeneity can robustly enhance random search processes even in a disordered
setting. This robustness could eventually be exploited in search strategies,
because it requires less prior knowledge about the location of the target.

\section{Global mean first passage time in a random system}

\begin{figure*}
\includegraphics[width=18.cm]{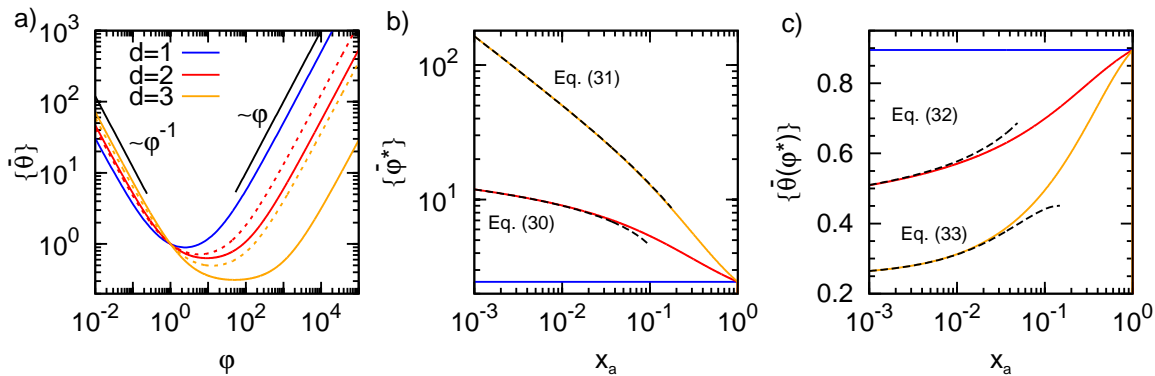}
\caption{a) Ratio $\bm{\{}\overline{\theta}\bm{\}}=\bm{\{}\overline{\mathbf{T}}
\bm{\}}/\overline{\mathrm{\mathbf{T}}}$ as a function of $\varphi=D_1/D_2$ for
different dimensions and  target size $x_a=0.1$ (full) and $x_a=0.01$ (dashed).
b) Optimal heterogeneity as a function of the target size. c) $\bm{\{}\overline{
\theta}\bm{\}}$ for the optimal strategy as a function of the target size.}
\label{FGGMFPT}
\end{figure*}

This section completes our study, addressing the Global MFPT in a random system, 
\begin{equation}
\bm{\{}\overline{\mathbf{T}}\bm{\}}=\frac{1}{1-x_a}\int_{x_a}^{1}\overline{
{\mathrm{\mathbf{T}}}}dx_{\mathrm{I}}.
\end{equation}
As far as blind spatially heterogeneous search is concerned this is the most robust
setting. It can be shown that the exact results for $\bm{\{}\overline{\mathbf{T}}
\bm{\}}$ have the form
\begin{widetext}
\begin{subequations}
\label{AP2X} 
\begin{eqnarray}
\label{GGMFPT1}
_1\bm{\{}\overline{\mathbf{T}}\bm{\}}&=&\frac{R^2}{60\overline{D}}\left[13+\varphi
+\frac{6}{\varphi}\right],\\[0.2cm]
\nonumber
_2\bm{\{}\overline{\mathbf{T}}\bm{\}}&=&\frac{R^2}{2\overline{D}(1-x_a^2)^2(1-x_a)}
\left[-\frac{127}{63\varphi}+\frac{293}{630}+\left(\frac{4}{3\varphi}-3\right)x_a+
\frac{1}{30}\left(\frac{40}{\varphi}+97-32\varphi\right)x_a^2\right.\\
\nonumber
&&+\frac{7}{18}\left(1-\frac{4}{\varphi}\right)\left(1+2\varphi\right)x_a^3-\frac{
1}{6}\left(\frac{2}{\varphi}+13\right)x_a^4+\frac{1}{30}\left(\frac{20}{\varphi}+
47+8\varphi\right)x_a^5+ \frac{1}{2}x_a^6\\[0.2cm]
\label{GGMFPT2}
&&\left.-\frac{1}{240}\left(\frac{20}{\varphi}+79+6\varphi\right)x_a^7-\frac{2}{3}
\left(\frac{2}{\varphi}+1+x_a\left[-3+x_a\{-3+x_a(1+2\varphi)\}\right]\right)\log
(x_a)\right],\\
\nonumber
_3\bm{\{}\overline{\mathbf{T}}\bm{\}}&=&\frac{R^2}{40\overline{D}(1-x_a^3)^2(1-x_a)}
\left[\frac{45}{x_a}\left(1+\frac{3}{\varphi}\right)-\frac{148}{\varphi}-361+5
\varphi+324x_a+135\left(\frac{1}{\varphi}-1\right)x_a^2\right.\\
\nonumber
&&+3\left(-\frac{45}{\varphi}+83+70\varphi\right)x_a^3-81(1+3\varphi)x_a^4-27\left(
\frac{1}{\varphi}+7\right)x_a^5+3\left(\frac{15}{\varphi}+47+10\varphi\right)x_a^6\\
\label{GGMFPT3}
&&\left.+ 36x_a^8-\left(\frac{5}{\varphi}+29+2\varphi \right)x_a^9+18\left(1-\frac{
1}{\varphi}\right)\left(x_a^3-\frac{1}{\varphi}\right)\log(x_a)\right].
\end{eqnarray}
\end{subequations}
The relative gain
\begin{equation}
\bm{\{}\overline{\theta}\bm{\}}=\frac{\bm{\{}\overline{\mathbf{T}}\bm{\}}}{
\overline{\mathbf{T}}}
\end{equation}
with respect to the Global MFPT for the standard random walk is shown in
Fig.~\ref{FGGMFPT}. The limiting scaling as $\varphi\to 0$ and $\varphi\to
\infty$ remains unchanged and the dependence of the overall gain on the target
size becomes significant for large $\varphi$.  The optimal heterogeneity in
the different spatial dimensions is obtained in the form
\begin{subequations} 
\begin{eqnarray}
\label{optiGG1}
_1\bm{\{}\overline{\varphi}\bm{\}}^{\ast}&=&\sqrt{6},\\
\label{optiGG2}
_2\bm{\{}\overline{\varphi}\bm{\}}^{\ast}&=&\sqrt{5}\left(\frac{-127-84\log(x_a)
+x_a(126+x_a[84-x_a\{98+3x_a(7-2x_a[7-x_a^2])\}])}{16-420x_a^3\log(x_a)-x_a^2(336
-x_a[245+3x_a^2\{28-3x_a^2\}])}\right)^{1/2},\\
\label{optiGG3}
_3\bm{\{}\overline{\varphi}\bm{\}}^{\ast}&=&\left(\frac{135/x_a-148+x_a^2(135-x_a
[135-x_a^2\{9-x_a^3\}])+180\log(x_a)}{5+x_a^3(210-x_a[243-x_a^2\{30-2x_a^3\}])+
180x_a^3\log(x_a)}\right)^{1/2}.
\end{eqnarray} 
\end{subequations}
\end{widetext}
The results are plotted in Fig.~\ref{FGGMFPT}a).  It can be shown that the
results in the $d=2$ and $d=3$ cases converge to the the result obtained in
$d=1$ in the limit $x_a\to 1$. This is expected since the system effectively
becomes one dimensional when the ratio annulus thickness-to-curvature
approaches zero. In the other limit $x_a\to 0$ we find the diverging optimal
heterogeneity (see Fig.~\ref{FGGMFPT}b)):
\begin{subequations}
\begin{equation}
\label{o2as}
_2\bm{\{}\overline{\varphi}\bm{\}}^{\ast}\simeq\frac{\sqrt{5[-127-84\log(x_a)]}}{4},
\end{equation}
in $d=2$, and
\begin{equation}
\label{o3as}
_3\bm{\{}\overline{\varphi}\bm{\}}^{\ast}\simeq \sqrt{\frac{27}{x_a}-\frac{148}{5}}
\end{equation}
\end{subequations}
in $d=3$. The gain of the optimal heterogeneity is shown in Fig.~\ref{FGGMFPT}c).
In $d=1$ the optimal gain is
\begin{subequations}
\begin{equation}
_1\bm{\{}\overline{\theta}\bm{\}}^{\ast}=(13+2\sqrt{6})/29\simeq 0.89495,
\end{equation}
and the expected convergence to this result in higher $d$ as $x_a\to 1$
is depicted in Fig.~\ref{FGGMFPT}c). In the corresponding limit $x_a\to 0$
the optimal gain in $d=2$ behaves as
\begin{widetext}
\begin{equation} 
\label{GGas}
_2\bm{\{}\overline{\theta}\bm{\}}^{\ast}\simeq\frac{\displaystyle{\frac{2032}{
\sqrt{5[-127-84\log(x_a)]}}}-293/5+\log(x_a)\left[\displaystyle{\frac{1344}{
\sqrt{5[-127-84\log(x_a)]}}}+84\right]}{63[3+4\log(x_a)]},
\end{equation}
\end{widetext}
ultimately converging to $1/3$ logarithmically slowly. Conversely, in $d=3$ the
gain converges to $1/4$ much more rapidly,
\begin{equation} 
\label{GGas3}
_3\bm{\{}\overline{\theta}\bm{\}}^{\ast}\simeq\frac{1}{4}+\sqrt{\frac{x_a}{12}}
-\left(\frac{47}{36}+\log(x_a)\right)x_a.
\end{equation} 
\end{subequations}
We therefore find that for vanishingly small targets an optimal heterogeneous
search in a random system configuration with uniformly distributed
starting position is remarkably 3 and 4 times faster, respectively in two
and three dimensions. We should stress here that this gain is achieved
under the constraint of a conserved $\overline{D}$, which means that we
do not introduce any additional resources. This is in striking contrast
to the optimization of active-passive intermittent strategies, where the
ballistic excursions based on active motion are by definition more expensive
\cite{oRMP,Bressloff,Mich1,Gleb1,Gleb2,ONatPh,LovPRE,BressNew1,BressNew2}. If
we were to relax the constraint on the conserved $\overline{D}$, the gain
could, in principle, become arbitrarily large.

\section{Discussion and concluding remarks}

We analyzed the kinetics of Brownian search in quenched heterogeneous
media. Analyzing a minimal model system, which captures all the essential
physical aspects of the problem, we obtained exact analytical results for
the MFPT of a particle to find the target in various settings. We showed that
the MFPT for Brownian search, both homogeneous and spatially heterogeneous,
is dominated by direct trajectories. Under the constraint of conserved
average dynamics, we proved the existence of an optimal heterogeneity,
which minimizes the MFPT.

We demonstrated and explained how a blind diffusive searcher in a spatially
heterogeneous environment can significantly outperform the homogeneous random
walk when the motion is faster near the target. This gain, which depends on
the size of the target, is significant and persists upon averaging over the
starting position, interface position, or even both. The enhancement is hence
very robust. In contrast to conventional search strategies (intermittent or
L\'evy-stable motion), which have the highest gain in lower dimensions, the
heterogeneous search performs best in higher dimensions. Because of the fact
that the MFPT is dominated by direct trajectories and the heterogeneity does
not affect the compactness of exploring the surrounding space---but instead
acts by enhancing or retarding the local dynamics---it performs better for
non-compact exploration.

According to recent single particle tracking experiments in living cells the
diffusivity of smaller proteins is faster close to the nucleus and slower
in the cell periphery \cite{het1,het2}. In addition, even in the presence
of quenched spatially disordered heterogeneity, which is often observed in
particle tracking experiments inside cells \cite{het3,het4,het2,Mhet2,Mhet3},
the target search kinetics can be enhanced as well, even if the process starts
from a spatially uniform initial distribution of the searching molecules.
Inside cells signaling proteins are found at extremely low concentrations,
down to, for instance, a dozen of $\lambda$-repressor molecules searching
for a single target in E. coli cell, whose volume is about 1 $\mu$m$^3$. The
search kinetics is thus central and rate-limiting for signaling dynamics.
Hence, heterogeneous search processes are important and relevant phenomena
at the few-molecule level.

At this point a few additional remarks are in order. The additivity in
Eq.~(\ref{MFPT}) only holds for the MFPT and not for higher moments. Based on
our present arguments the additivity principle in Eq.~(\ref{MFPT}) should hold
for an arbitrary number of segments but a formal proof is part of our current
investigation. If this is indeed the case, an optimal heterogeneity function
can be formally constructed for any situation by first taking the appropriate
limits of small segments and afterward optimized using variational methods. The
underlying physical principle however, will remain unchanged. This will allow
the study of more realistic heterogeneity profiles such as those observed
experimentally in living cells \cite{het1,het2,het4,het3,Mhet2,Mhet3}.

In the present context the optimization of diffusion heterogeneity
cannot be perceived as a search strategy in the traditional sense
\cite{oRMP,Bressloff,Holcman,Sheinmann,Max,Otto,Mich3}. A searcher would
somehow have to know the position of the target in order to optimize his
motion pattern according to a optimal heterogeneity. Conversely, in a cell
the position of the target is given and so are the cytoplasm and nucleus
properties giving rise to a spatially varying diffusivity.  However, it could
be part of an evolutionary optimization to improve the search efficiency of
biomolecules in cell regulatory processes.

The present results can be extended and generalized in numerous ways, the
immediate extension being the study of the full distribution of first passage
times.  Furthermore, within the context of a traditional search strategy our
findings cannot be directly used to asses the efficiency of a heterogeneous
search with a general off-center target position as in \cite{OlRaph}. The
optimization of the target position-averaged MFPT with respect to the
ratio $\varphi$ of diffusivities is part of an ongoing investigation. Given
the present results we can speculate, however, that an enhancement might
indeed be possible in terms of an optimal heterogeneity at least for target
positions not too close to the external boundary. One can therefore imagine
that heterogeneous search strategies are also beneficial for computer search
or stochastic minimization algorithms. The ideas can be generalized to
diffusion processes in more complex disordered systems \cite{dis1,dis2,Sinai}
and even anomalous diffusion processes of continuous time random walk type
\cite{Johannes}. Namely, in a finite system the heavy-tailed waiting time
density between individual jumps in a subdiffusive continuous time random
walk is expected to be exponentially tempered, exhibiting subdiffusion over a
transient but long time scale, which would ultimately terminate with a normal
diffusion regime. In such a system the MFPT to the target will be finite and
it would be interesting to investigate whether and how  the existence and
properties of an {\em optimal heterogeneity\/} change in a heterogeneous
system with transiently subdiffusive dynamics. The most obvious extension
of the present results, however, goes in to the direction of heterogeneous
intermittent search. Clearly, a combination of both could lead to a highly
superior search dynamics

\begin{acknowledgments}
The authors thank Andrey Cherstvy, Olivier B{\'e}nichou and Raphael Voituriez
for stimulating discussions and Sidney Redner for suggestions and critical
reading of the manuscript. AG acknowledges funding through an Alexander von
Humboldt Fellowship. RM acknowledges funding from the Academy of Finland
(FiDiPro scheme).
\end{acknowledgments}

\appendix

\section{Computer simulation results}

\begin{figure*}
\includegraphics[width=18.cm]{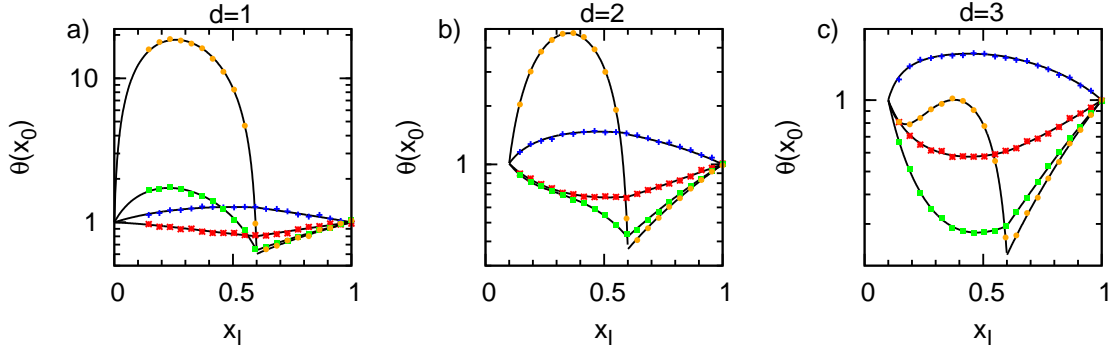}
\caption{Ratio $\theta=\mathrm{\mathbf{T}}(x_0)/\mathrm{\mathbf{T}}^0(x_0)$
as function of $x_{\mathrm{I}}$ for $x_0=0.6$ and $x_a=0.1$ in different
dimensions. The colors depict results for various heterogeneities:
$\varphi=$0.6 (blue), 2 (red), 10 (green), and 150 (orange). The symbols
are results of simulations and the full lines correspond to the analytical
results in Eqs. (\ref{GNRL}).}
\label{AMFPT}
\end{figure*}

\begin{figure*}
\includegraphics[width=18.cm]{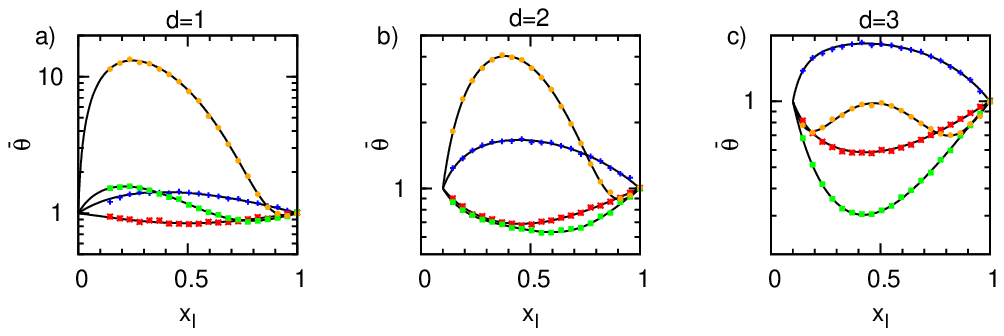}
\caption{Ratio
$\overline{\theta}=\overline{\mathbf{T}}/\overline{\mathbf{T}}^0$ as function
of $x_{\mathrm{I}}$ for $x_a=0.1$ in different dimensions. The colors
depict results for various heterogeneities: $\varphi=$0.5 (blue), 2 (red),
10 (green), and 120 (orange). The symbols are results of simulations and
the full lines correspond to the analytical results in Eqs. (\ref{AP2X}).}
\label{AGMFPT}
\end{figure*}

\begin{figure*}
\includegraphics[width=18.cm]{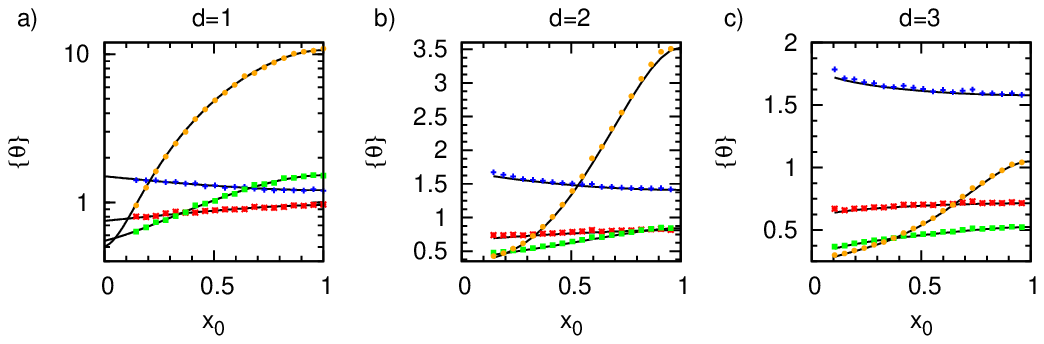}
\caption{Ratio $\bm{\{}{\theta}\bm{\}}=\bm{\{}\mathbf{T}\bm{\}}/\mathbf{T}(x_0)$
as function of $x_{\mathrm{0}}$ for $x_a=0.1$ in different dimensions. The
colors depict results for various heterogeneities: $\varphi=$0.5 (blue), 2
(red), 10 (green), and 120 (orange). The symbols are results of simulations
and the full lines correspond to the analytical results in Eqs. (\ref{AP1X}).}
\label{AIMFPT}
\end{figure*}

\begin{figure*}
\includegraphics[width=18.cm]{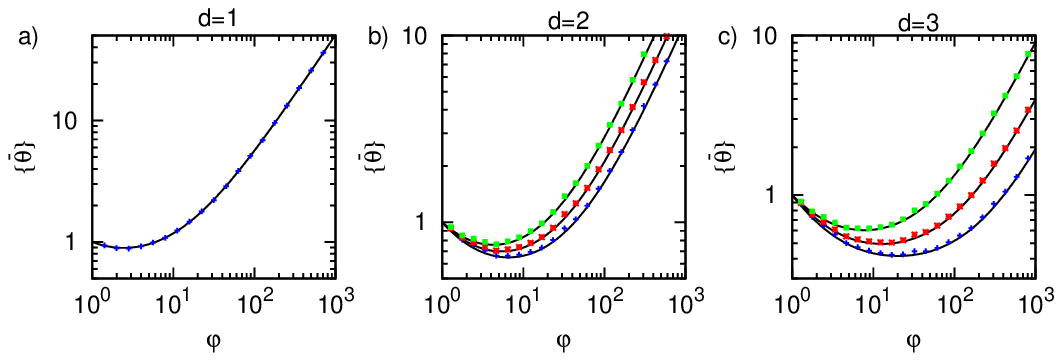}
\caption{a) Ratio $\bm{\{}\overline{\theta}\bm{\}}=\bm{\{}\overline{\mathbf{T}}
\bm{\}}/\overline{\mathrm{\mathbf{T}}}$ as a function of $\varphi$ for
different dimensions and target sizes. In dimensions 2 (b) and 3 (c) the target
sizes correspond to 0.05 (blue), 0.1 (red) and 0.2 (green), respectively.}
\label{AGGMFPT}
\end{figure*}

Computer simulations were performed according to the scheme of transition
probabilities outlined in Sec. II. 10$^5$ trajectories were simulated for
each set of $x_a$, $x_i$, and/or $x_0$. The simulations correspond to a
discrete random walk in between spherical shells with an absorbing boundary
at $x=x_a$ and a reflecting one at $x=1$ recording the number of steps within
each region $n_{1,2}$ as well as the total number of steps $n$ to obtain the
process time according to
\begin{equation}
t = \frac{\Delta R^2}{2D_1}(n_1-n_2+n)+\frac{\Delta R^2}{2D_2}(n_2-n_1+n).
\tag{A1}
\end{equation} 
Note that $n$ also counts the number of steps in the interfacial shells.
The results for various cases are plotted below. We find a remarkably good
agreement between our analytical and the simulation results. The MFPT
as a function of the interface position is shown in Fig.~\ref{AMFPT}.
Intuitively, the MFPT depends strongly on the starting and interface
positions and can exhibit none, one, or two local minima as a function
of $x_{\mathrm{I}}$. Hence, it is possible for a given $\varphi$ that two
distinct interface positions lead to the same MFPT. It should be noted that
an equal $\varphi$ does not correspond to equal $D_1$ and $D_2$, only their
ratio is fixed. A degeneracy of the MFPT depending on the interface position
is thus not surprising. The simulation results for the Global MFPT are shown
in Fig.~\ref{AGMFPT} and compared to the analytical predictions of Section V.
In the case of the Global MFPT as well, zero, one or two minima are observed
depending on $\varphi$. Here we observe significant differences in the features
of the Global MFTP at equal $\varphi$ in different dimension. This is due to
the fact that the statistical weight of various starting positions is different
in different dimensions and more distant starting configurations have a higher
weight in higher dimensions. The simulation results for the interface position
averaged MFPT in a random system are shown in Fig.~\ref{AIMFPT} and compared to
the analytical predictions of Section VI.  In the case of MFPT with a random
interface position, the disorder averaged MFPT (Fig.~\ref{AIMFPT}) is only
weakly dependent on $x_0$ as long as $\varphi$ is not too large. Again, very
distinct behavior is found for different dimensions at equal $\varphi$. In
contrast to the Global MFPT this originates solely from the differences in
the exploration of space. The results for the interface position averaged
Global MFPT in a random system are shown in Fig.~\ref{AGGMFPT}.  Also here
we find an excellent agreement between theory and simulation. We see that
the dependence on the target size becomes more prominent in higher dimensions
because of the different statistical weight of starting positions at a given
distance from  $x_a$ and and the differences in sampling space.


\begin{thebibliography}{99}
\bibitem{REM} More precisely, the spatially heterogeneous dissipation has a purely
fluctuational origin and does not reflect heterogeneities in the static free volume
accessible to the proteins---and would hence give rise to a spatially uniform
steady-state profile in absence of the absorbing target.
\end{thebibliography}
\end{document}